\documentclass[11pt,twoside]{article}
\usepackage[left=3cm,right=3cm,top=3cm,bottom=3cm]{geometry}
\usepackage{amssymb}
\usepackage{graphicx}
\usepackage{graphicx,type1cm,eso-pic,color}
\usepackage{amssymb}
\usepackage{amssymb,amsmath}
\usepackage{graphicx,epsfig}
\usepackage{enumerate}
\usepackage{color}
\usepackage {multicol}
\usepackage{lineno}
\numberwithin{equation}{section}
\newtheorem{theorem}{Theorem}
\newtheorem{lemma}{Lemma}

\newtheorem{definition}{Definition}
\newtheorem{proposition}{Proposition}

\newtheorem{example}{Example}

\makeatletter
\AddToShipoutPicture{
	\setlength{\@tempdimb}{.5\paperwidth}
	\setlength{\@tempdimc}{.5\paperheight}
	\setlength{\unitlength}{1pt}
	\put(\strip@pt\@tempdimb,\strip@pt\@tempdimc)
}
\makeatother

\usepackage{scalerel,stackengine}
\stackMath
\newcommand\reallywidehat[1]{%
	\savestack{\tmpbox}{\stretchto{%
			\scaleto{%
				\scalerel*[\widthof{\ensuremath{#1}}]{\kern-.6pt\bigwedge\kern-.6pt}%
				{\rule[-\textheight/2]{1ex}{\textheight}}
			}{\textheight}%
		}{0.5ex}}%
	\stackon[1pt]{#1}{\tmpbox}%
}
\begin{document}
	\setcounter{page}{1}
	\thispagestyle{empty}
	\markboth{}{}

	\pagestyle{myheadings}
	\markboth{}{ }
	
	\pagestyle{myheadings}
	\markboth{Chaudhary et al.}{ Chaudhary et al.}
	
	\date{}
	
	
	\noindent  
	
	\vspace{.1in}
	
	{\baselineskip 20truept

		\begin{center}
			{\Large {\bf On cumulative past information
					generating function}} \footnote{\noindent	{\bf 1.}  Corresponding author E-mail: skchaudhary1994@gmail.com\\
				{\bf 2. } E-mail: nitin.gupta@maths.iitkgp.ac.in \\
				{\bf 3. } E-mail: acroy21@gmail.com }
	\end{center}}

	\vspace{.1in}
	
	\begin{center}
		{\large {\bf Santosh Kumar Chaudhary$^1$, Nitin Gupta$^2,$ Achintya Roy$^3$}}\\
		\vspace{.1in}
		{\large {\it ${}^{1,2}$ Department of Mathematics, Indian Institute of Technology Kharagpur, West Bengal 721302, India. }} \\
		{\large {\it ${}^{1}$	Centre for Quantitative Economics and Data Science, Birla Institute of Technology Mesra, Ranchi, Jharkhand 835215, India. }}\\			
		{\large {\it ${}^{3}$ Department of Mathematics and Basic Sciences, NIIT University, Neemrana, Rajasthan 301705, India. }}\\

	\end{center}
	
	\vspace{.1in}
	\baselineskip 12truept

	
	\begin{center}
		{\bf \large Abstract}\\
	\end{center}
	In this paper, we study the cumulative past information-generating function (CPIG) and relative cumulative past information-generating function (RCPIG). We study its properties. We establish its relation with generalized cumulative past entropy (GCPE). We defined the CPIG stochastic order and its relation with dispersive order. We provide the results for the CPIG measure of the convoluted random variables in terms of the measures of its components. We found some inequality relating to Shannon entropy, CPIG and GCPE. Some characterization and estimation results are also discussed regarding CPIG.  We defined divergence measures between two random variables, Jensen-cumulative past information generating function(JCPIG), Jensen fractional cumulative past entropy measure, cumulative past Taneja entropy, and Jensen cumulative past Taneja entropy information measure.  \\
	\\
	\textbf{Keyword:} Cumulative past entropy, Cumulative past extropy, Cumulative past information generating function,  Entropy, Extropy, Relative cumulative past information generating function. \\
	\\
	\noindent  {\bf Mathematical Subject Classification}: {\it 62B10, 62D05}
	\section{Introduction}\label{s1intro}
	Information measures can be used to measure the amount of uncertainty about random experiments. Measuring the uncertainty associated with a random variable is crucial in many practical scenarios in real-life data analysis, experimental physics, econometrics, and demography. With Shannon's (1948) introduction of the Shannon entropy or differential entropy idea, the foundational work on information theory was created. This entropy has a wide range of applications in diverse areas such as financial analysis, data compression, molecular biology, hydrology, meteorology, computer science, and information theory.
	
	Let $X$ be a random variable with probability density  function (pdf) $f_X,$ distribution function $F_X(x)$ and survival function $\Bar{F}_X(x),\ x \in \mathbb{R},$ and
	let $l=inf\{x\in \mathbb{R}: F_X(x)>0\}$ and $r=sup\{x\in \mathbb{R}: \Bar{F}_X(x)>0\}$ respectively denote the lower and upper limits of support of $X.$ The differential entropy of an absolutely continuous random variable $X$ is given by	
	\begin{align}
		H(X)=-\int_{l}^{r}  f_X(x) \log( f_X(x))dx,
	\end{align}
	where $\log$ means natural logarithm and, by convention, $0 \log 0 =0.$
	
	In the literature, various entropy measures have been introduced, each of which is appropriate in a different set of circumstances. The cumulative residual entropy (CRE) which measures the uncertainty in the	future of a lifetime of a system, by Rao et al. (2004) is defined as 
	\begin{align}
		\xi(X)=-\int_{l}^{r}  \bar{F}_X(x) \log (\bar{F}_X(x))dx.
	\end{align}
	Di Crescenzo and Longobardi (2009) introduced cumulative entropy  for estimating the uncertainty in the past life of a system as cumulative past entropy (CPE) is defined as
	\begin{align*}
		\tilde \xi(X)= -\int_{l}^{r}  F_X(x) \log(F_X(x))dx.
	\end{align*}

	In recent years, there has been a great interest in the generalization of CRE and CPE. The generalized cumulative residual entropy (GCRE) of order $n$ proposed by Psarrakos and Navarro (2013), is defined as
	\[\xi_n(X)=\frac{1}{n!} \int_{l}^{r}  \bar{F}_X(x) (-\log(\bar{F}_X(x)))^ndx.\] 
	The generalized cumulative past entropy (GCPE) of order $n$ is defined as
	\[\tilde \xi_n(X)=\frac{1}{n!} \int_{l}^{r}  F_X(x) (-\log(F_X(x)))^ndx.\]
	Note that  when $n=1, \ \tilde \xi_n(X)=\tilde \xi(X)$ and $\xi_n(X)=\xi(X).$
	
	Golomb (1966) proposed an information-generating (IG) function as
	\begin{align*}
		IG_\alpha(X)= \int_{l}^{r}  f_X^\alpha(x)dx, \ \alpha>0.
	\end{align*}
	The relative information generating function, for any $\alpha>0$, is
	defined as
	\begin{align*}
		RIG_\alpha(X,Y)=\int_{l}^{r}  f_X^\alpha (x) f_Y^{1-\alpha} (x) dx.
	\end{align*}
	The derivatives of this IG function with respect to $\alpha$ at $\alpha=1$ 	yield statistical information measures for a probability distribution. Note that the first-order derivative of IG function with respect to $\alpha$ at $\alpha=1$	produces a negative of Shannon entropy measure. Kharazmi and Balakrishnan (2021a) considered the IG function and discussed some new properties that reveal its connections to some other well-known information measures. They have shown that the IG function can be expressed based on different orders of fractional Shannon entropy. Kharazmi et al. (2021) studied IG function and relative IG function measures associated with maximum and minimum ranked set sampling schemes with unequal sizes. Kharazmi and Balakrishnan (2021b) considered the cumulative residual information generating function (CPIG) and relative cumulative residual information generating function (RCPIG) and discussed some new properties that reveal their connections to some other well-known information measures. Kharazmi and Balakrishnan (2021b) considered the CRIG function and relative CRIG respectively, for $\theta>0,$ as
	\begin{align}
		CRIG_\theta(X)&=	-\frac{1}{2} \int_{l}^{r}  \bar{F}_X^\theta(x)dx,  \\
		RCRIG_\theta(X,Y)&=\int_{l}^{r}  \bar{F}_X^\theta (x) \bar{F}_Y^\theta (x) dx,
	\end{align}    
	and discussed some new properties that reveal its connections to some other well-known information measures. Capaldo et al. (2023) introduced and studied the cumulative information-generating function (CIGF), which provides a unifying mathematical tool suitable to deal with classical and fractional	entropies based on the cumulative distribution function and the survival function.  Chakraborty and Pradhan (2024) considered the cumulative residual information generating (CRIG) function and studied some new properties. The existing IG functions, the relative IG function, the CRIG function and the relative CRIG function provided us motivation for considering the cumulative past information generating function (CPIG) and relative cumulative past information generating function (RCPIG), which are discussed in detail in the subsequent sections of this paper. 
	
	Extropy, an alternative measure of uncertainty, was defined by Lad et al. (2015) and proved to be the complement dual of the Shannon entropy. The extropy also called the differential extropy of a absolutely continuous random variable with probability density function $f_X(x)$ is 
	\[J(X)=-\frac{1}{2}\int_{l}^{r} f_X^2(x)dx.\]	
	Cumulative past extropy (CPJ) and cumulative residual extropy (CRJ) are respectively, given as
	\begin{align*}
		CPJ(X)=\bar{\xi}J(X)=-\frac{1}{2} \int_{l}^{r}  F_X^2(x)dx, \\
		CRJ(X)=\xi J(X)=-\frac{1}{2} \int_{l}^{r}  \bar{F}_X^2(x)dx.
	\end{align*}
	
	The rest of this paper is organized as follows. In section \ref{s2oncpig}, we consider the CPIG and establish some new properties that reveal its connections to some other well-known information measures. It is also
	shown that the CPIG can be expressed based on different orders of generalized cumulative past entropy. Some results associated with stochastic	ordering are also provided in section \ref{s3Stochticorder}. The convolution-related results are studied in section \ref{s4convolution}. Some lower bounds to CPIJ are provided in section \ref{s5inequlities}. In section \ref{s6characterization}, we provide a few more characterizations. Section \ref{s7estimation} is devoted to empirical estimations and related properties.  The Jensen cumulative past-information generating function is proposed in section \ref{s8JCPIGF}. Finally, section \ref{s9conclusion} presents some concluding remarks.

	\section{On cumulative past information generating function}\label{s2oncpig} 
	Capaldo et al. (2023) defined CIGF for continuous random variables $X$ as 
	\begin{align}\label{CIGF}
		G_X(\alpha, \beta)= \int_{l}^{r} F_X^{\alpha}(x) \Bar{F}_X^{\beta}(x)dx.
	\end{align}         
	
	When we take $\alpha=0$ in ($\ref{CIGF}$) then CIGF becomes a CRIG function. CRIG function and relative CRIG function have been defined by Kharazmi and Balakrishnan (2021b). Chakraborty and Pradhan (2024) studied the CRIG function in detail. When we take $\beta=0$ in ($\ref{CIGF}$) then CIGF becomes a cumulative past information generating (CPIG) function. We will study CPIG and relative CPIG in this paper.

	Let $X$ be a continuous rv with cdf $F$, the CPIG function is given as
	\begin{align}\label{defCPIG}
		CPIG_\theta(X)=	\tilde \zeta_\theta(X) &=\int_{l}^{r} F_X^\theta(x)dx,\ \theta>0.
	\end{align}
	The definition of relative CPIG is given below.
	\begin{definition}
		Let $X$ and $Y$ be two continuous random variables with absolutely continuous cdf $F_X$ and $f_Y$, respectively. Then, the relative cumulative past information generating (RCPIG) measure between $X$ and $Y$ (or $F_X$	and $F_Y$), for any $\theta>0$,	is defined as
		\begin{align}\label{defRCPIG}
			RCPIG_\theta(X,Y)=\int_{l}^{r} F_X^\theta (x) G_Y^\theta (x) dx.
		\end{align}
	\end{definition}
	
	Let us consider some propositions regarding CPIG measures that relate CPIG to other information measures.
	\begin{proposition}\label{cpigcpj}
		Suppose the random variable $X$ has cdf $F$. Then, a new representation of the CPIG measure is given by 
		
		\[  \tilde \zeta_2(X)=-2 \tilde{\xi}J(X). \]
	\end{proposition}
	\textbf{Proof}  Proof follows easily from the definition of $\tilde \zeta_2(X)$ and $\tilde{\xi}J(X).$
	
	\begin{proposition}
		Suppose the random variable $X$ has cdf $F$. Then, a new representation of the CPIG measure is given by 		
		\[ \frac{\partial}{\partial \theta}\tilde \zeta_\theta(X)|_{\theta =1}=-\tilde \xi(X).  \]
	\end{proposition}
	\textbf{Proof} From (\ref{defCPIG}),
	\[	\frac{\partial}{\partial \theta}\tilde \zeta_\theta(X)=\int_{0}^{\infty}F_X^\theta(x) \log(F_X(x))dx.\]
	Hence, the result follows.

	\begin{proposition}
		Suppose the random variable $X$ has cdf $F$. Then, a new representation of the CPIG measure is given by 		
		\[ \frac{\partial^n}{\partial \theta^n}\tilde \zeta_\theta(X)|_{\theta =1}=(-1)^n n!\tilde \xi_n(X). \]
	\end{proposition}
	\textbf{Proof} Using the principle of mathematical induction, the result follows.

	The gini mean difference (GMD) of random variable $X,$ $GMD(X)$ is  defined as 
	\begin{align*}
		GMD({X})=2 \int_{l}^{r} \bar{F}_X(x)(1-\bar{F}_X(x))dx
	\end{align*}	
	\begin{proposition}
		Suppose the random variable $X$ has cdf $F$. Then, a new representation of the CPIG measure is given by 		
		\[\tilde \zeta_1(X)-\tilde \zeta_2(X)= \frac{GMD({X})}{2}.\]	
	\end{proposition}
	\textbf{Proof} Taking differece of $\tilde \zeta_\theta(X)$ after putting $\theta=1$ and $\theta=2$ in (\ref{defCPIG}), proof follows.
	
	Now we see the following preposition which shows the relationship 
	of $\tilde \zeta_\theta(X) $ with other measures.	
	\begin{theorem}
		Suppose the random variable $X$ has cdf $F$. Then, a new representation of the CPIG measure is given by 
		
		\[\tilde \zeta_\theta(X)=\sum_{n=0}^{\infty} (1-\theta)^n \tilde{\xi}_n(X). \]
	\end{theorem}
	\textbf{Proof}
	Using Maclaurin series expansion, Fubini's theorem and definition of CPIG measure as in (\ref{defCPIG}), we have 
	\begin{align*}
		\tilde \zeta_\theta(X)&=\int_{l}^{r}F^\theta(x)dx\\
		&=\int_{l}^{r}e^{(\theta-1) \log(F(x))}F(x)dx \\
		&=\int_{l}^{r}\left(\sum_{n=0}^{\infty} \frac{((\theta-1) \log(F(x)))^n}{n!} \right)F(x)dx \\
		&=\sum_{n=0}^{\infty} (1-\theta)^n \int_{l}^{r}\frac{(-\log(F(x)))^n F(x)}{n!}dx\\
		&= \sum_{n=0}^{\infty} (1-\theta)^n \tilde{\xi}_n(X).
	\end{align*}

	The following proposition provides some properties of $\tilde \zeta (T)$ in relatioship with CPIG of $n$th order statistics	$X_{(n)}.$
	\begin{theorem}
		Consider a random variable $X$ with CPIG measure $\tilde \zeta (X)$, then
		
		$\tilde \zeta_\theta (X_{(n)})\leq \tilde \zeta_\theta (X).$

	\end{theorem}
	\textbf{Proof} Using definition of CPIG and cdf of $X_{(n)}$, 
	\[\tilde{\zeta}_\theta (X_{(n)})=\int_{l}^{r} F^\theta_{(n)}(v)dv=\int_{l}^{r} F^{n\theta}(v)dv \leq \int_{l}^{r} F^\theta(v)dv=\tilde{\zeta}(X).\]
	\begin{theorem}
		Consider a random variable $X$ with CPIG measure $\tilde \zeta (X)$, then
		$X_1\leq_{st} X_2$, implies that $\tilde \zeta (X_1)\geq \tilde \zeta (X_2)$.
	\end{theorem}
	\textbf{Proof} From definition of stochastic ordering, $X_1\leq_{st} X_2$ implies $ F_1(v) \geq F_2(v).$ Hence, $$\tilde{\zeta}(X_1)\geq \tilde{\zeta}(X_2).$$

	\begin{example}
		Let $X$ be $U(a,b)$ random variable where $b>a>0,$ then $\tilde \zeta_\theta(X)=\frac{b-a}{\theta+1}.$
	\end{example}
	
	\begin{example}
		Let $X$ be $U(a,b)$ random variable where $b>a>0,$ then using Proposition \ref{cpigcpj}, we obatain $ \tilde{\xi}J(X) =- \frac{b-a}{6}.$
	\end{example}

	\section{CPIJ stochastic order}\label{s3Stochticorder}
	
	\begin{definition}(Shaked and Shanthikumar (2007))
		Suppose X and Y are two continuous random variables with distributions $F_X$ and $F_Y$ and density functions $f_X$ and $f_Y$, respectively. Then, we say that X is less dispersed than Y (denoted by $X \leq_{disp} Y$) if $f_X(F_X^{-1}(t))\geq f_Y(F_Y^{-1}(t))$ for all $t\in(0,1).$ 
	\end{definition}

	\begin{definition} 
		Let $X$ and $Y$ be two variables with CPIG measures $\tilde \zeta_\theta(X)$ and $\tilde \zeta_\theta(Y)$  respectively, as defined in (\ref{defCPIG}). Then, $X$ is said to be less than $Y$ in cumulative past information generating function measure, denoted by $X\leq_{CPIG}Y,$ if $\tilde \zeta_\theta(X)\leq \tilde \zeta_\theta(Y).$
		
	\end{definition}
	
	\begin{theorem}
		Suppose $X \leq_{disp} Y.$ Then, for $\theta>0, \ X \leq_{CPIG} Y.$
	\end{theorem}
	\textbf{Proof} From (\ref{defCPIG}), we can write
	\begin{align*}
		\tilde \zeta_\theta(X)=\int_{l}^{r} F_X^\theta (x)dx=\int_{0}^{1} \frac{t^\theta}{f_X(F_X^{-1}(t))} dt, \\
		\text{and}\ \ \tilde \zeta_\theta(Y)=\int_{l}^{r} F_Y^\theta (t)dt=\int_{0}^{1} \frac{t^\theta}{f_Y(F_Y^{-1}(t))} dt.
	\end{align*}
	Since  $X \leq_{disp} Y,$ which implies $\tilde \zeta_\theta(X) \leq \tilde \zeta_\theta(Y), $ hence, $X \leq_{CPIG} Y.$

	Let X be a random variable with a continuous cdf $F_X$, and $\phi$ be an increasing, differentiable and invertible function. Let $W=\phi(X)$ then we have 
	\begin{align*}
		\tilde \zeta_\theta(W)=\int_{l}^{r} F_W^\theta (w)dw= \int_{c}^{d} F_X^\theta (x) \phi^{\prime}(x)dx,
	\end{align*}
	where $c=\phi^{-1}(0)$ and $d=\phi^{-1}(\infty).$

	\begin{theorem}
		Suppose $X\leq_{disp}Y.$ If $\phi$ is convex and increasing, then $\phi(X)\leq_{CPIG}\phi(Y).$
	\end{theorem}
	\textbf{Proof} From (\ref{defCPIG}) and under the conditions $\phi^{-1}(0)=0$ and $\phi^{-1}(\infty)=\infty,$ we have
	\begin{align*}
		\tilde \zeta_\theta(\phi(X))= \int_{l}^{r}  F_X^\theta (x) \phi^{\prime}(x)dx= \int_{0}^{1} \frac{u^\theta \phi^{\prime}(F_X^{-1}(u))}{f(F_X^{-1}(u))}du,\\
		\text{and} \ \ \tilde \zeta_\theta(\phi(Y))= \int_{l}^{r}  F_Y^\theta (y) \phi^{\prime}(y)dy=\int_{0}^{1} \frac{u^\theta \phi^{\prime}(F_Y^{-1}(u))}{f(F_Y^{-1}(u))}du. 
	\end{align*}
	Since $X \leq_{disp} Y,$ therefore $f_X(F_X^{-1}(t))\geq f_Y(F_Y^{-1}(t))$ for  all $t\in(0,1).$ As $\phi(x)$ is convex and increasing, so $\phi^\prime(x)$ is increasing and positive. Since  $X \leq_{disp} Y$ implies $ X \leq_{st} Y$ (See, Jeon et al. (2006)). This means $\phi^\prime(F_X^{-1}(t))\leq \phi^\prime(F_Y^{-1}(t)) $ for  all $t\in(0,1)$ and hence,
	\begin{align*}
		\tilde \zeta_\theta(\phi(X))=\int_{0}^{1} \frac{u^\theta \phi^{\prime}(F_X^{-1}(u))}{f_X(F_X^{-1}(u))}du \leq \int_{0}^{1} \frac{u^\theta \phi^{\prime}(F_Y^{-1}(u))}{f_Y(F_Y^{-1}(u))}du=\tilde \zeta_\theta(\phi(Y)).
	\end{align*}

	\section{ Convolution }\label{s4convolution}
	
	We now take into account the convolution of random variables and provide the results for the CPIG measure of the convoluted variable in terms of the measures of its components. First, using CPIG measures for each of its components, we derive an upper bound for the CPIG measure of the convolution of two independent random variables. We will expand it to $n$ independent random variables. We here noted down that CPE can be generated from the CPIG function. We also expressed CPIG as a weighted sum of GCPE of order $n$. Now here in the following theorem we study the CPIG function for the convolution of two independent random variables.
	
	\begin{theorem}\label{convolution1}
		Let $X$ and $Y$ be two independent random variables with cdf $F$   and $G,$  respectively. Then, for $\theta \geq (\leq) 1, \ \tilde \zeta_\theta(X+Y)\leq (\geq) min\{\tilde \zeta_\theta(X), \tilde \zeta_\theta(Y)\}.$ 
	\end{theorem}
	\textbf{Proof}
	The cdf of the convoluted random variable $X+Y$ is given by 
	\[F_{X+Y}(t)=\int_{l}^{r} F_X(t-y)dF_Y(y).\]	
	Using Jensen’s inequality, we have, for $\theta \geq 1,$
	\[F^\theta_{X+Y}(t) \leq \int_{l}^{r} F^\theta_X(t-y)dF_Y(y).\]
	Now, using definition (\ref{defCPIG}) and Jensen’s inequality, we obtain
	\begin{align*}
		\tilde \zeta_\theta(X+Y) &=\int_{l}^{r} F^\theta_{X+Y}(t)dt \leq \int_{l}^{r} \left(\int_{l}^{r} F^\theta_X(t-y)dF_Y(y)\right) dt \\
		&=\int_{l}^{r} \left(\int_{y}^{\infty} F^\theta_X(t-y) dt\right) dF_Y(y) \\
		&=\int_{l}^{r} \left(\int_{l}^{r} F^\theta_X(u) du \right) dF_Y(y)  \\
		&= \left(\int_{l}^{r} F^\theta_X(u) du \right) \left(\int_{l}^{r} dF_Y(y) \right) \\
		&= \int_{l}^{r} F^\theta_X(u) du\\
		&= \tilde \zeta_\theta(X).
	\end{align*}	
	Thus, $\tilde \zeta_\theta(X+Y) \leq \tilde \zeta_\theta (X).$ Similar to this, we can show that $\tilde \zeta_\theta(X+Y) \leq \tilde \zeta_\theta (Y).$ Therefore,  $\tilde \zeta_\theta(X+Y)\leq min\{\tilde \zeta_\theta(X), \tilde \zeta_\theta(Y)\}.$ Hence the result.

	Theorem \ref{convolution1} can easily be expanded to the convolution of $n$ independent random variables as shown in the following.
	\begin{theorem}
		Let $X_1,\dots,X_n$ be independent random variables with cdf $F_1,\dots, F_n,$ respectively. Then, for $\theta \geq (\leq) 1,$ \[ \tilde \zeta_\theta \left(\sum_{i=1}^{n} X_i \right) \leq (\geq) min\left(\tilde \zeta_\theta(X_1), \dots,  \tilde \zeta_\theta(X_n)\right).\] 
	\end{theorem}
	\textbf{Proof}
	By using Theorem \ref{convolution1}, we immediately have $$\tilde \zeta_\theta \left(\sum_{i=1}^{n} X_i \right) \leq min\left(\tilde \zeta_\theta \left(\sum_{i=1}^{n-1} X_i\right), \tilde \zeta_\theta(X_n)\right).$$ 
	Then, by using this inequality recursively, we obtain\\
	$$\tilde \zeta_\theta \left(\sum_{i=1}^{n} X_i \right) \leq min\left(\tilde \zeta_\theta(X_1), \dots,  \tilde \zeta_\theta(X_n)\right).$$

	\section{Inequalities}\label{s5inequlities}
	We here study some bounds for $\tilde \zeta (X)$ in terms of Shannon entropy and CPE.
	\begin{theorem}
		Consider $X$ to be a continuous rv, and then the following inequalities hold.
		\begin{enumerate}[(i)]
			\item $ \tilde \zeta (X)\geq e^{H(X)-\theta},$
			
			\item $ \tilde \zeta (X)\geq  \left(\int_{l}^{r} F_X(v)dv\right) \  \exp{\left(-\dfrac{(\theta-1)\tilde \xi(X)}{\int_{l}^{r} F_X(v)dv}\right)}$ \text{where}  $\exp(x)=e^x.$
		\end{enumerate}
		
	\end{theorem} 
	\textbf{Proof}
	\begin{enumerate}
		\item[(i)] Using log-sum inequality,
		\begin{align*}
			\int_{l}^{r} f_X(v)\log\left(\frac{f_X(v)}{F_X^\theta(v)}\right)dv \geq \left(\int_{l}^{r} f_X(v)dv\right) \log\left(\frac{\int_{l}^{r} f_X(v)dv}{\int_{l}^{r} F_X^\theta (v)dv}\right)=-\log(\tilde \zeta (X)).
		\end{align*}
		Further simplification implies \[ \tilde \zeta (X)\geq e^{H(X)-\theta}.\]
		\item[(ii)] 
		\begin{align*}
			\int_{l}^{r} F_X(v)\log\left(\frac{F_X(v)}{F_X^\theta(v)}\right)dv \geq \left(\int_{l}^{r} F_X(v)dv\right) \log\left(\frac{\int_{l}^{r} F_X(v)dv}{\int_{l}^{r} F_X^\theta (v)dv}\right).
		\end{align*}
		Further simplification implies
		\[ \tilde \zeta (X)\geq  \left(\int_{l}^{r} F_X(v)dv\right) \  \exp{\left(-\dfrac{(\theta-1)\tilde \xi(X)}{\int_{l}^{r} F_X(v)dv}\right)}.\]
	\end{enumerate}

	\begin{theorem}
		Consider a rv $X$ with CPIG measure $\tilde \zeta_{\theta} (X)$, then for $\theta>1$,
		\[\tilde \zeta_{\theta} (X)\geq \left(\frac{\theta-1}{\theta}\right)^{\theta}\int_{l}^{r}\left(\frac{1}{v}\int_{0}^{v}F_X(t)dt\right)^{\theta}dv.\]
	\end{theorem}
	\textbf{Proof}
	From Hardy's inequality (see Hardy (1920), Walker (2015), Chakraborty and Pradhan (2023)),
	
	\[\int_{l}^{r} \left( \frac{1}{v}\int_{0}^{v}F_X(v)dv\right)^\theta dv \leq \left(\frac{\theta}{\theta-1}\right)^\theta \int_{l}^{r} F_X^\theta (v)dv, \ \theta > 1. \]
	Further simplification implies	
	\[\tilde \zeta_{\theta} (X)\geq \left(\frac{\theta-1}{\theta}\right)^{\theta}\int_{l}^{r}\left(\frac{1}{v}\int_{0}^{v}F_X(t)dt\right)^{\theta}dv, \ \  \theta > 1.\]

	\section{Characterizations}\label{s6characterization}
	
	\begin{lemma}\label{lemma1sec2}
		For any sequence of positive integers $\{n_j:j\geq1\},$ the sequence of polynomials $\{X^{n_j}\}$ is complete on L(0,1) iff $\sum_{j=1}^{\infty}n_{j}^{-1}=\infty.$
	\end{lemma}
	Let $X_{(1)},X_{(2)},\ldots,X_{(n)}$ be the order statistics with respect to the random sample $X_1,X_2,\ldots,X_n$. Let the cdf of $X_{(n)}$ be $F_{(n)}$. Then the CPIG measure for $X_{(n)}$ is
	\[\tilde \zeta (X_{(n)})=\int_{l}^{r} F_{(n)}^{\theta}(v)dv=\int_{l}^{r}\left(F_X^n(v)\right)^{\theta}dv=\int_{0}^{1}\frac{u^{n\theta}}{f_X\left(F_X^{-1}(u)\right)}du,\ u\in (0,1).\]

	\begin{theorem}
		Let $X$ and $Y$ be two continuous rvs with pdfs $f_X,\  f_Y$ and cdfs $F_X,\  F_Y$, respectively. Then $X$ and $Y$ belong to the same location family of distribution iff
		$\tilde \zeta (X_{(n)})=\tilde \zeta (Y_{(n)}),\ \forall \  n=n_j,\ j\geq 1,$ such that $\sum_{j=1}^{\infty}n_{j}^{-1}=\infty.$
	\end{theorem}	
	\textbf{Proof} If $X=Y+a$ then $F_X(x)=F_Y(x-a)$ and hence, $\tilde \zeta (X_{(n)})=\tilde \zeta (Y_{(n)}).$	Conversely, suppose $\tilde \zeta (X_{(n)})=\tilde \zeta (Y_{(n)}),$ then 
	\begin{equation}\label{assumption1}
		\int_{l}^{r} w^{n\theta} \left[ \frac{1}{f_X(F_X(w))}-\frac{1}{f_Y(F_Y(w))} \right]dw=0.
	\end{equation}
	If (\ref{assumption1}) holds for all $n=n_j,\ j\ \geq \ 1,$ such that  $\sum_{k=1}^{\infty}n_{j}^{-1}=\infty,$ then from Lemma \ref{lemma1sec2}, we have $f_X(F_X(w))=f_Y(F_Y(w)).$ Since $\frac{d}{dw} F_1^{-1}(w)=\frac{1}{f_X(F_X^{-1}(w))},$ we get $F_X^{-1}(w)=F_Y^{-1}(w)+constant.$ Therefore, $F_X$ and $F_Y$ belong to the same family of distribution but for a change of location.

	\begin{theorem}
		Let $X$ and $Y$ be two rvs with common support $(0,\infty)$ and $E(X_{(n)})$ and $E(Y_{(n)})$ exist. Then $X$ and $Y$ belong to the same location family of distribution iff
		$\frac{\tilde \zeta (X_{(n)})}{E(X_{(n)})}=\frac{\tilde \zeta (Y_{(n)})}{E(Y_{(n)})},\ \forall n=n_j,\ j\geq 1,$ such that $\sum_{k=1}^{\infty}n_{k}^{-1}=\infty.$
	\end{theorem}
	\textbf{Proof}
	The proof is similar to that of Theorem 2 of Abbasnejad (2011).

	\section{Estimation}\label{s7estimation}
	Here in this section, a non-parametric estimate of the CPIG function is obtained and we study its properties. The estimator is obtained using edf. Consider $X_1,X_2,\ldots,X_n$ be a random sample  drawn from a continuous cdf $F$ and $X_{(1)}\leq X_{(2)} \leq X_{(3)} \leq \dots \leq X_{(n)}$ denote order statistics of random sample $X_1, X_2, \dots, X_n.$ The empirical distribution function of $F$ is defined as 		
	\begin{eqnarray*}
		{F}_{n}(v)=
		\begin{cases}
			0, \hspace{4mm} v<X_{(1)}\\
			\frac{i}{n}, \hspace{4mm} X_{(i)}\leq v<X_{(i+1)}, \ \ \ i=1,2,\dots, n-1,\\
			1, \hspace{5mm} v \geq X_{(n)}.
		\end{cases}
	\end{eqnarray*}
	The estimator of $\tilde \zeta_\theta(X)$ is 
	\begin{align}
		\tilde \zeta_\theta(F_n) &=\int_{l}^{r}F_n^\theta(t)dt \nonumber\\
		&= \sum_{t=0}^{n-1} \int_{X_{(i)}}^{X_{(i+1)}} \left(\frac{i}{n}\right)^\theta \nonumber\\
		&= \sum_{t=0}^{n-1} \left(X_{(i+1)}-X_{(i)}\right) \left(\frac{i}{n}\right)^\theta,
	\end{align}
	where $T_{(0)}=0.$ 
	
	\begin{theorem}
		Let $X\in L^p, \ p>1$ then $\tilde \zeta_\theta(F_n)$ converges almost surely to   $\tilde \zeta_\theta(X).$
	\end{theorem}
	\textbf{Proof}
	The proof follows similar to that of Theorem 9 of Rao et al (2004).

	\begin{example}
		Let a random sample $X_1, X_2, \dots, X_n$ of size $n$ is taken from exponential distribution with mean $\frac{1}{\lambda}.$ The sample spacing $X_{(i+1)}- X_{(i)}, \ i=1,2,3, \dots, n-1$ exponential distribution with mean $\frac{1}{\lambda(n-i)}$ (see Pyke (1965)). The mean and variance of $\tilde \zeta_\theta(X)$ is obtained as 
		\[\mathbb{E}(\tilde \zeta_\theta(F_n))= \frac{1}{\lambda}\sum_{t=0}^{n-1} \frac{
			1}{n-i} \left(\frac{i}{n}\right)^\theta\]
		and 
		\[Var(\tilde \zeta_\theta(F_n))= \frac{1}{\lambda^2}\sum_{t=0}^{n-1} \frac{
			1}{(n-i)^2} \left(\frac{i}{n}\right)^\theta.\]
	\end{example}
	\begin{theorem}
		For a random sample $X_1, X_2, \dots, X_n$ of size $n$ from exponential distribution with mean $\frac{1}{\lambda},$
		\[\frac{\tilde \zeta_\theta(F_n)-\mathbb{E}(\tilde \zeta_\theta(F_n))}{(Var(\tilde \zeta_\theta(F_n)))^{\frac{1}{2}}} \overset{d}{\longrightarrow} N(0,1) \] as $n \rightarrow \infty.$
	\end{theorem}
	\textbf{Proof}
	The proof is similar to that of Theorem 7.1 of Di-Crescenzo and Longobardi(2009).

	\begin{example}
		Let a random sample $X_1, X_2, \dots, X_n$ of size $n$ is taken from uniform  distribution on interval $(0,1)$. The sample spacing $T_{(i+1)}- T_{(i)}, \ i=1,2,3, \dots, n-1$ beta distribution (1,n)(see Pyke (1965)). The mean and variance of $\tilde \zeta_\theta(T)$ is obtained as 
		\[\mathbb{E}(\tilde \zeta_\theta(F_n))=\frac{1}{n+1} \sum_{t=0}^{n-1}  \left(\frac{i}{n}\right)^\theta\]
		and 
		\[Var(\tilde \zeta_\theta(F_n))=\frac{n}{(n+1)^2(n+2)} \sum_{t=0}^{n-1}  \left(\frac{i}{n}\right)^\theta. \]
	\end{example}

	\section{Jensen-cumulative past information generating function}\label{s8JCPIGF}
	\begin{definition}\label{defDFxFy}
		Let $F_X$ and $F_Y$ be two cdf of non-nagative random variables $X$ and $Y$, respectively, with respective CPIG measures $\tilde \zeta_\theta(X)$ and $\tilde \zeta_\theta(Y).$ Then, a divergence measure between $F_X$ and $F_Y,$ $D_\theta(F_X, F_Y),$ is defined as 
		\begin{align*}
			D_\theta(F_X,F_Y)=  
			\begin{cases}
				\int_{l}^{r} F^\theta_X(x) L_{\frac{1}{\theta}} \left(\frac{F^\theta_X(x)}{F^\theta_Y(x)}\right)dx - (\tilde \zeta_\theta(X)-\tilde \zeta_\theta(Y)),&  \theta\geq 1\\
				(\tilde \zeta_\theta(X)-\tilde \zeta_\theta(Y))-\int_{l}^{r} F^\theta_X(x) L_{\frac{1}{\theta}} \left(\frac{F^\theta_X(x)}{F^\theta_Y(x)}\right)dx,              & 0< \theta\leq 1
			\end{cases} 
		\end{align*}
		where $L_q(z)$ is defined as 	
		\begin{align*}
			L_q(z)=
			\begin{cases}
				\frac{z^{1-q}-1}{1-q},\ & z>0, \ q\in [0, 1)\cup(1, \infty)\\
				\log z,\  & z > 0, \ q = 1,
			\end{cases}
		\end{align*}
		and known as the generalized logarithm function with $L_q(z)\rightarrow \log z$ as $q\rightarrow1$ (See Asadi et al.(2017)). In a similar manner, we can define $D_\theta(F_Y, F_X).$
	\end{definition}
	
	\begin{theorem}
		Let $F_X$ and $F_Y$ be two cdf with CPIG measures $\tilde \zeta_\theta (X)$ and $\tilde \zeta_\theta (Y)$ , respectively. Then, $D_\theta(F_X,F_Y)$ is non-negative.
	\end{theorem}
	\textbf{Proof} From Lemma 2.1 of Asadi et al. (2017), the desired result naturally follows.
	
	\begin{definition}\label{defJCPIGFXFY}
		Let $X$ and $Y$ be random  variables with cdf $F_X$ and $F_Y,$ respectively. Then, the Jensen-cumulative past information generating (JCPIG) function, for any $0<p<1,$ is defined as
		
		\begin{align*}
			JCPIG_\theta(F_X, F_Y; p)=
			\begin{cases}
				\tilde \zeta_\theta(pF_X+(1-p)F_Y)-(p\tilde \zeta_\theta(F_X)+(1-p)\tilde \zeta_\theta(F_Y)), &0<\theta\leq 1, \\
				p\tilde \zeta_\theta(F_X)+(1-p)\tilde \zeta_\theta(F_Y)-\tilde \zeta_\theta(pF_X+(1-p)F_Y), &\theta \geq 1.
			\end{cases}
		\end{align*}
	\end{definition}
	
	\begin{theorem}
		Let random variables $X$ and $Y$ have cdf $F_X$ and $F_Y$,  respectively. Then, the Jensen-cumulative past information generating function, $JCPIG_\theta(F_X, F_Y; p),$ for $\theta>0,$ is a mixture of the measure of the form
		\begin{align*}
			JCPIG_\theta(F_X, F_Y; p)=pD_\theta(F_X,F_T)+(1-p)D_\theta(F_Y,F_T),
		\end{align*}
		where $D_\theta(F_X, F_T)$ and  $D_\theta(F_Y, F_T)$ are divergence measure given in Definition \ref{defDFxFy}, and $F_T=pF_X+(1-p)F_Y$ is mised cdf based on $F_X$ and $F_Y.$
	\end{theorem}
	\textbf{Proof} Proof is similar to the proof of Theorem 5.4 in Kharazami and Balakrishnan (2021b).
	
	The Definition \ref{defJCPIGFXFY} can be generalized to the case of more than two populations as follows.
	\begin{definition}
		Let $X_1, X_2, \dots X_n$ be random  variables with cdf $F_1, F_2,\dots, F_n$ respectively. Then, the Jensen-cumulative past information generating (JCPIG) function, for any $0<p<1,$ is defined as
		\begin{align*}
			JCPIG_\theta(F_1, F_2,\dots, F_n; p)=
			\begin{cases}
				\tilde \zeta_\theta(\sum_{i=1}^{n}p_iF_i)-\sum_{i=1}^{n} p_i \tilde \zeta_\theta(F_i), &0<\theta\leq 1, \\
				\sum_{i=1}^{n} p_i \tilde \zeta_\theta(F_i)-\tilde \zeta_\theta(\sum_{i=1}^{n}p_iF_i), &\theta \geq 1.
			\end{cases}
		\end{align*}
	\end{definition}
	
	\begin{theorem}
		The  Jensen-cumulative  past  information  generating  function $JCPIG_\theta(F_1, F_2,\dots, F_n; \textbf{p})$ is a mixture of the form
		\[JCPIG_\theta(F_1, F_2,\dots, F_n; \textbf{p})=\sum_{i=1}^{n} p_i D_\theta(F_i, F_T),\]           
		where $D_\theta$ is the divergence measure defined in Definition \ref{defDFxFy},  $F_T=\sum_{i=1}^{n}p_iF_i $ is the mixed cdf based  on  the components $F_1, F_2, \dots, F_n$ and $\textbf{p}=(p_1, p_2, p_3, \dots, p_n), \ p_i \geq 0$ such that  $\sum_{i=1}^{n} p_i=1.$     
	\end{theorem}
	\textbf{Proof} Proof is similar to the proof of Theorem 5.5 in Kharazami and Balakrishnan (2021b).
	
	\begin{definition}( Jensen-fractional cumulative past entropy measure) 
		Let $X_1, X_2, \dots, X_n$ be random variables with cdfs $F_1, F_2, \dots, F_n,$ respectively, and $p_1, p_2, \dots, p_n$ be non-negative real numbers such that $\sum_{i=1}^{n} p_i=1.$ Then, the Jensen-fractional cumulative past entropy (JFCPE) information measure is defined as 
		\begin{align*}
			JFCPE_q(F_1, F_2,\dots, F_n; \textbf{p})=\bar{\phi}_q \left(\sum_{i=1}^{n} p_i F_i\right)-\sum_{i=1}^{n}p_i \bar{\phi}_q(F_i)
		\end{align*}	
		where $\bar{\phi}_q(F)=\int_{l}^{r} F(x)(-\log F(x))^q$ is fractional cumulative past entropy of order $q$ and $q\in(0,1).$
	\end{definition}
	
	\begin{theorem}
		The JFCPE information measure,$JFCPE_q(F_1, F_2,\dots, F_n; \textbf{p})$, is non-negative.
	\end{theorem}
	\textbf{Proof} Proof follows from the fact that the function $x (-\log(x))^q$ is concave for $0<x<1$ and $0<q\leq 1.$ 
	
	Next, we define a cumulative version of Taneja entropy using cdf in place of the density function; for more details about Taneja entropy, see Sharma and Taneja(1975).

	\begin{definition}
		If X is a variable with cdf $F,$ then the cumulative past Taneja entropy (CPTE) of order $q>1$ is defined as
		\begin{align*}
			CPTE_q(F)=-2^{q-1} \int_{l}^{r} F^q(x) \log(F(x))dx.
		\end{align*}
	\end{definition}
	
	\begin{definition}
		Let $X_1, X_2, \dots, X_n$ be random variables with cdf $F_1, F_2, \dots, F_n$, respectively, and $\alpha_1, \alpha_2, \dots, \alpha_n$ be non-negative real numbers such that $\sum_{i=1}^{n} \alpha_i=1.$ Then, the Jensen-cumulative past Taneja entropy (JCPTE) information measure is defined as
		\begin{align*}
			JCPTE_q(F_1, F_2, \dots, F_n; \textbf{n})= CPTE_q \left(\sum_{i=1}^{n} p_i F_i\right)-\sum_{i=1}^{n}p_i CPTE_q(F_i).
		\end{align*}
	\end{definition}
	\begin{theorem}
		The JCRTE information measure, $JCPTE_q(F_1, F_2, \dots, F_n; \textbf{n})$, is non-negative.
	\end{theorem}
	\textbf{Proof} Proof follows from the fact that the function $-x^q \log(x)$ is concave for $0<x<1$ and $0<q\leq 1.$

	\section{Conclusion}\label{s9conclusion}
	In this research, we studied cumulative past information generating (CPIG) and relative cumulative past information generating (RCPIG) measures. The CPIG measurement gives more information about cumulative entropic measures.  We also provided related characterization, inequalities, stochastic order, convolution and estimation. Finally, we have defined the Jensen-cumulative past information generating function, Jensen-cumulative past  Taneja entropy,  Jensen-fractional cumulative past entropy and Jensen-cumulative past information measure. In future work, we will study more about the Jensen-cumulative past information generating function, Jensen-cumulative past  Taneja entropy,  Jensen-fractional cumulative past entropy and Jensen-cumulative past information measure.\\
	\\	
	\noindent \textbf{ \Large Conflict of interest} \\
	\\
	No conflicts of interest are disclosed by the authors.\\
	\\
	\textbf{ \Large Funding} \\
	\\
	Santosh Kumar Chaudhary would like to acknowledge financial support from the Council of Scientific and Industrial Research (CSIR) ( File Number 09/0081(14002) /2022-EMR-I ), Government of India. \\
		

\begin{thebibliography}{99}
		
		\bibitem{citekey} Capaldo, M., Di Crescenzo, A. \& Meoli, A., 2023. Cumulative information generating function and generalized Gini functions. Metrika. https://doi.org/10.1007/s00184-023-00931-3
		
		\bibitem{citekey} Asadi, M., Ebrahimi, N. and Soofi, E.S., 2017. Connections of Gini, Fisher, and Shannon by Bayes risk under proportional hazards. Journal of Applied Probability, 54(4), pp.1027-1050.		
		
		\bibitem{citekey} Asadi, M., Ebrahimi, N., Soofi, E.S. and Zohrevand, Y., 2016. Jensen–Shannon information of the coherent system lifetime. Reliability Engineering \& System Safety, 156, pp.244-255.
		
		\bibitem{citekey} Chakraborty, S., Pradhan, B., 2024. On cumulative residual information generating function: properties, inference and applications. OPSEARCH. https://doi.org/10.1007/s12597-024-00754-4.
		
		\bibitem{citekey} Di Crescenzo, A. and Longobardi, M., 2009. On cumulative entropies. Journal of Statistical Planning and Inference, 139(12), pp.4072-4087.
		
		\bibitem{citekey} Hardy, G.H., 1920. Note on a theorem of Hilbert. Mathematische Zeitschrift, 6(3-4), pp.314-317.
		
		\bibitem{citekey} Jeon, J., Kochar, S. and Park, C.G., 2006. Dispersive ordering—some applications and examples. Statistical Papers, 47, pp.227-247.
		
		
		\bibitem{citekey} Kharazmi, O. and Balakrishnan, N., 2021a. Jensen-information generating function and its connections to some well-known information measures. Statistics \& Probability Letters, 170(1), p.108995.
		
		\bibitem{citekey} Kharazmi, O. and Balakrishnan, N., 2021b. Cumulative and relative cumulative residual information generating measures and associated properties. Communications in Statistics-Theory and Methods, 52(15), pp.5260-5273.
		
		\bibitem{citekey} Kharazmi, O., Tamandi, M. and Balakrishnan, N., 2021. Information Generating Function of Ranked Set Samples. Entropy, 23(11), p.1381.
		
		\bibitem{citekey} Lad, F., Sanfilippo, G. and Agro, G., 2015. Extropy: Complementary dual of entropy, 30(1), pp.40-58.
		
		\bibitem{citekey} Psarrakos, G. and Navarro, J., 2013. Generalized cumulative residual entropy and record values. Metrika, 76(1), pp.623-640.
		
		
		\bibitem{citekey} Pyke, R., 1965. Spacings. Journal of the Royal Statistical Society: Series B (Methodological), 27(3), pp.395-436.
		
		
		\bibitem{citekey} Rao, M., Chen, Y., Vemuri, B.C. and Wang, F., 2004. Cumulative residual entropy: a new measure of information. IEEE Transactions on Information Theory, 50(6), pp.1220-1228.
		
		
		\bibitem{citekey} Shannon, C.E., 1948. A mathematical theory of communication. The Bell System Technical Journal, 27(3), pp.379-423.
		
		\bibitem{citekey} Sharma, B.D. and Taneja, I.J., 1975. Entropy of type $(\alpha, \beta)$ and other generalized measures in information theory. Metrika, 22(1), pp.205-215.
		
		\bibitem{citekey} Shaked, M. and Shanthikumar, J.G. eds., 2007. Stochastic orders. New York, NY: Springer New York.
		
		\bibitem{citekey} Walker, S.G., 2015. A probabilistic proof of the Hardy inequality. Statistics \& Probability Letters, 103, pp.6-7. 
		
		
		\bibitem{citekey}	Kumar, V. \& Taneja, H.C. (2012). On dynamic cumulative inaccuracy measure. In Proceedings of the World Congress on Engineering, 4–6 July 2012, London, UK.
		
		\bibitem{citekey}	Di Crescenzo, A., Longobardi, M., 2009. On cumulative entropies. Journal of Statistical Planning and
		Inference. 139, 4072-4087.
		
		\bibitem{citekey}	Kerridge, D.F., 1961. Inaccuracy and inference. J. R. Stat. Soc. Ser. B Stat. Methodol. 23 (1), 184-194.
		
	\end{thebibliography}
\end{document}